%Paper: gr-qc/9508013
%From: MIGNEMI@VAXCA.CA.INFN.IT
%Date: Sat, 5 Aug 1995 17:30:54 +0300 (CET-DST)

\def\c{\chi}\def\d{\delta}\def\e{\epsilon}
\def\g{\gamma}
\def\m{\mu}\def\n{\nu}\def\o{\omega}\def\q{\psi}\def\r{\rho}
\def\y{\eta}\def\x{\xi}

\def\G{\Gamma}\def\L{\Lambda}
\def\O{\Omega}\def\Q{\Psi}

\def\lie{{\cal L}}\def\de{\partial}\def\na{\nabla}
\def\inf{\infty}\def\id{\equiv}\def\ha{{1\over 2}}
\def\qu{{1\over 4}}

\def\const{{\rm const}}

\def\ep{\e_{\m\n}}\def\mn{{\mu\nu}}

\def\tors#1#2#3{T_{#1#2#3}}\def\curv#1#2#3#4{R_{#1#2#3#4}}
\def\af{asymptotically flat }\def\hd{higher derivative }
\def\bh{black hole }
\def\tran{transformations }

\def\RC{Riemann-Cartan }\def\poi{Poincar\'e }

\def\section#1{\bigskip\noindent{\bf#1}\smallskip}

\def\PL#1{Phys.\ Lett.\ {\bf#1}}\def\CMP#1{Commun.\ Math.\ Phys.\ {\bf#1}}
\def\PRL#1{Phys.\ Rev.\ Lett.\ {\bf#1}}\def\AP#1#2{Ann.\ Phys.\ (#1) {\bf#2}}
\def\PR#1{Phys.\ Rev.\ {\bf#1}}\def\CQG#1{Class.\ Quantum Grav.\ {\bf#1}}

\def\JMP#1{J.\ Math.\ Phys.\ {\bf#1}}\def\PTP#1{Prog.\ Theor.\ Phys.\ {\bf#1}}

 \def\IJMP#1{Int.\ J. Mod.\ Phys.\ {\bf #1}}
\def\MPL#1{Mod.\ Phys.\ Lett.\ {\bf #1}}

\def\ref#1{\medskip\everypar={\hangindent 2\parindent}#1}
\def\beginref{\begingroup
\bigskip
\centerline{\bf References}
\nobreak\noindent}
\def\endref{\par\endgroup}
\def\ab{_{ab}}\def\erp{e^{2\r}}
\def\hm{{h-1}}\def\hp{{h+1}}\def\Lq{{\L\over 4}}\def\gi{{1\over\g}}
\def\Lh{{\L\over 2}}\def\gh{{\g\over 2}}\def\edx{\int e\ d^2x}
\def\xo{$x_1$}\def\xt{$x_2$}\def\conn#1{\o\ba_{\ \  #1}}
\def\abm{{^a_{\ b}}}\def\bam{{_a^{\ b}}}\def\ba{^{ab}}

\magnification=1200
{\nopagenumbers
June 1994\hfill IRS-9402
\vskip120pt
\centerline{\bf Exact solutions, symmetries and quantization}
\centerline{\bf of two-dimensional higher-derivative gravity with
dynamical torsion}
\vskip40pt
\centerline{\bf S. Mignemi\footnote{$^\dagger$}{\rm
e-mail: mignemi@cagliari.infn.it}}
\vskip20pt
\centerline{Dipartimento di Scienze Fisiche, Universit\`a di Cagliari}
\centerline{via Ospedale 72, 09100 Cagliari, Italy}
\vskip80pt
\centerline{\bf ABSTRACT}
{\noindent We investigate two-dimensional \hd gravitational theories in a
\RC framework. We obtain the most general static
black hole solutions in conformal coordinates and discuss their geometry.
We also consider the hamiltonian formulation of the theory and discuss
its symmetries, showing that it can be
considered as a gauge theory of a non-linear generalizations of the
2-dimensional \poi algebra. We also show that the models can be exactly
quantized in the Dirac formalism.}
\vfil\eject}

\section{1. Introduction}
In this paper, we study a general class of two-dimensional higher-derivative
gravitational theories in a Riemann-Cartan spacetime. As is well known, in two
dimensions the Einstein-Hilbert action is a total derivative and hence cannot
be used to construct a two-dimensional version of general relativity. It is
however possible to construct actions whose lagrangian density is given by
an arbitrary power of the Ricci scalar [1]. These models turn out to be
equivalent to the ordinary Einstein-Hilbert action non-minimally coupled to
a scalar field with a power-law potential [2-4]. In such formalism the field
equations become second order and it is possible to obtain exact solutions
and to perform the Dirac quantization of the theory [5]. Some well-known
special cases of gravity-scalar theories in two dimensions
include the Jackiw-Teitelboim [6] and the "string" models [7-8].

Another useful generalization of two-dimensional gravity is given by the
consideration of \RC geometries with non-trivial torsion. Several authors
have studied various aspects of a model with action quadratic in
the curvature and the torsion [9-15]. It has been proved that this model is
completely integrable [9-10] and, exploiting the hamiltonian formalism, it has
been shown that its symmetries are a realization of a specific non-linear
algebra [11-12] and that its quantization can be performed exactly [13].

Finally, a further interesting aspect of some two-dimensional models is that
they can be interpreted as gauge theories
of the \poi or de Sitter group in two dimensions [16]. In ref. [14] it has
been shown that this interpretation can be extended to theories with
non-trivial torsion, provided that one generalizes the notion of gauge
invariance
to groups generated by non-linear algebras.

In this paper we try to extend the results obtained so far to the case of
actions
containing arbitrary powers of the curvature and quadratic torsion.
We show that the static solutions of the field equations can be found exactly
and discuss their geometry. We also investigate the hamiltonian formulation
of the models and find the constraint algebra, which is a non-linear
deformation of the two-dimensional \poi algebra and therefore permits their
interpretation as a non-linear gauge theory of such algebra.
Finally, we perform the Dirac quantization of the
model and define the space of the physical states.

\section{2. The action and the field equations}

We consider a 2-dimensional lorentzian manifold with signature $(-,+)$
endowed with a \RC geometry.
The geometry can be described by the zweibein field $e^a_\m$  and the Lorentz
connection $\conn{\m}$, where $a,b,..$ are tangent space indices which can
take the values $0$, $1$ and $\m,\n,...$ are world indices, whose values will
be denoted by $t$, $x$.

The curvature and torsion are defined as
$$\eqalign{&R\ba_\mn=\de_\m\conn{\n}-\de_\n\conn{\m}\cr
&T^a_\mn=\de_\m e^a_\n-\de_\n e^a_\m+\o\abm_\m e^b_\n-\o\abm_\n e^b_\m\cr}
\eqno(1)$$
and the Ricci scalar as $R=e_a^\m e_b^\n R\ba_\mn$. In two
dimensions the Ricci scalar determines uniquely the Riemann tensor by
the relation $\curv abcd=-\ha\e_{ab}\e_{cd}R$.
Moreover, the connection can be written as
$$\conn{\m}=\e\ba\o_\m$$
where $\e\ba$ is the antisymmetric tensor $\e\ab=-\e\ba$, $\e_{01}=1$.

In this paper, we study higher derivative actions of the form:
$$S=\edx (R^k-\gh T^2)\eqno(2)$$
where $e=$det $e^a_\m$, $T^2=\tors abc T^{abc}$.
Here, $k$ is any real number except $0$, $1$ and $\g$ a coupling constant.
These actions generalize to \RC geometry the higher-derivative actions
introduced in [1] and further investigated in [2-5]. The special case $k=2$ has
already been studied by several authors [9-15].

By defining a scalar field $\y=kR^{k-1}$, it is possible, by a standard
argument [2-4] to reduce the action (2) to a form which is linear in the
curvature:
$$S'=\edx\left(\y R-\gh T^2+\L\y^h\right)\eqno(3)$$
where $\L=(1-k)k^{-k/(k-1)}$ and $h={k\over k-1}$.

The action can be further reduced to a fully first order form, by introducing
a doublet of scalar fields $\y_a$ [13]:
$$S''=\edx\left(\y R+\y_a\, ^*T^a-\gi\y_a\y^a+\L\y^h\right)\eqno(4)$$
where $^*T^a=\e^{bc}\tors abc$.

The field equations obtained by varying (3) with respect to the fields
$\y$, $e^a$ and $\o$ can be written as:
$$\eqalign{&R+h\L\y^\hm=0\cr
&-\na^b\tors cdb +T\ba_{\ \ c}\tors abd-\qu g_{cd}(T^2-{2\L\over\g}\y^h)=0\cr
&\e^a_{\ c}e_a^\m\de_\m\y-\g\e\ba\tors abc=0\cr}\eqno(5)$$

\section{3. The static solutions}

In the following, we shall look for the static solutions of these equations.
According to the discussion of the special case $k=2$
performed in ref. [9], it results convenient
to seek for the solutions in a conformal gauge. We therefore adopt the ansatz:
$$e^0_t=e^1_x=e^{2\r(x)}\qquad\qquad e^0_x=e^1_t=0$$
We also assume that $\o_\m=\ep\de^\n\c$, with $\c=\c(x)$, which yields
$$\o_x=0,\qquad\qquad \o_t=\c'$$
where a prime denotes derivative with respect to $x$.
In terms of these variables, one has:
$$R=2e^{-2\r}\c'',\qquad\qquad\tors 001=e^{-\r}(\r'-\c')\eqno(6)$$
and the field equations become:
$$\c''+h\Lh\y^\hm\erp=0\eqno(7.a)$$
$$\y'-\g(\r'-\c')=0\eqno(7.b)$$
$$\r''-\c''-\ha(\r'-\c')(\r'+\c')+{\L\over 2\g}\y^h\erp=0\eqno(7.c)$$
$$\ha(\r'-\c')(\r'+\c')+{\L\over 2\g}\y^h\erp=0\eqno(7.d)$$

These equations admit a special solution $\c=\r=\y=0$,
corresponding to a manifold with vanishing
curvature and torsion (we are cosidering the case of vanishing cosmological
constant). The zero torsion solutions do not therefore reduce to the extremals
of the action (3) with $\g=0$, which have been discussed in [5]. This fact has
been observed in [9] for the special case $k=2$.

One can however obtain more general solutions to (7).
Combining (7.c) and (7.d) one gets
$$\r''-\c''=(\r'-\c')(\r'+\c')=-{\L\over\g}\y^h\erp\eqno(8)$$
A first integral of the first equation (8) is
$$\r=\ha(f+\ln(Ef'))\eqno(9)$$
where $f\equiv\r-\c$ and $E$ is an integration constant.

{}From (7.b), one has $\y=\g f$ and thus (8) yields
$$f''=\L E\g^\hm e^ff^hf'\eqno(10)$$
which can be integrated to give
$$f'=\L E\g^\hm\int^f_0 g^he^gdg\ +A\eqno(11)$$
where the integral on the r.h.s. is proportional to the incomplete gamma
function $\G(h+1,-f)$ and $A$ is an integration constant. The equation (11)
can then be integrated numerically
to give $f$. This solution generalizes that obtained in ref. [9] for $h=2$.

One can now express the curvature and the torsion in terms of the function
$f$:
$$R=-\L(\g f)^\hm\qquad\qquad T^2=e^{-f}f'\eqno(12)$$
It is then easy to study the qualitative behaviour of the solutions of (11).
We shall impose the positivity of $f$ in order to avoid problems when $h$ is
not integer. From the asymptotics of (11), follows that $f\to\inf$ for a finite
value of $x$. If $h>1$, a curvature singularity is therefore always present
at finite $x$ in these coordinates, while the scalar $T^2$ is singular if
$h>0$ as $f\to\inf$.

A numerical study permits to distinguish three different possible behaviours
for $f$ (see fig. 1-3):

a) If $h>-1$ and $A>0$, $f$ grows monotonically from $0$ to $\inf$ between
two finite values \xt\ and \xo\ of $x$.

b) If $h>-1$ and $A<0$, the solution has two branches: one of them decreases
monotonically between a constant value $f_0$ at $x=-\inf$ and 0 at $x=$ \xo,
while the other grows monotonically between $f_0$ at $-\inf$ and infinity at
$x=$\xt.

c) If $h<-1$, for any $A$, the behaviour of $f$ is similar to the case b),
but $f'(x_2)\to-\inf$.

In order to investigate the properties of the solutions near the critical
point, one must study the behavior of  the functions $R$, $T^2$ and $\erp$
near \xo\ and \xt. It is easy to check that for $f\to\inf$, $R\sim f^\hm$,
$T^2\sim f^h$, $\erp\sim e^{2f}$, while for $f\to 0$, $R\sim f^\hm$,
$T^2\sim\const +f^{h+1}$, $\erp\sim\const +f^{h+1}$.

Depending on the value of $h$, one can then distinguish several cases
which are summarized in the tables 1-3. The following general results can be
stated:
If $h>-1$ and $A>0$, a naked singularity  is always present, either at \xo\ or
at \xt.
More interesting are the cases where $h>-1$ and $A<0$ or $h>-1$. In these
cases, the two branches of the solution describe the interior and the exterior
of an \af\bh with the horizon located at $x=-\inf$. For $h<1$ (resp. $h>1$),
the curvature singularity is at \xo (resp. \xt), while spatial infinity is at
\xt\ (resp. \xo). For $0<h<1$, however, the torsion diverges at spatial
infinity.

Of course, a detailed study of the spacetime structure would require
a more explicit form of the solutions.
For the special cases $h=0$ and $h=1$, this  will be afforded in a future
paper.

\section{4. First order formalism}
In terms of differential forms, $e^a=e^a_\m dx^\m$, $\o=\o_\m dx^\m$,
the first order lagrangian in (4) can be written as:
$$\ha\lie = \y_2 d\o+\y_aT^a+\left(\Lq\y_2^h+{1\over 2\g}\y_c\y^c\right)\e\ab
e^ae^b
\eqno(13)$$
where we have renamed $\y$ as $\y_2$.

This form of the lagrangian
is especially convenient because it permits to evidentiate the
connection of our models with the formulation of 2-dimensional gravity as a
gauge theory of the Poincar\'e group ISO(1,1) or one of its generalizations
[16,8].
In this formalism, $e^a$ and $\o$ play the role of gauge connections, while
the $\y$ are considered as auxiliary fields.
Local Poincar\'e \tran with parameters $\x^2$ and $\x^a$, corresponding to
Lorentz rotations and to translations, act infinitesimally on the
fields according to:
$$\d e^a=d\x^a+\e\ab(\x^b\o-\x^2e^b)\qquad\qquad\d\o=d\x^2$$
$$\d\y_a=\e\bam\x^2\y_b\qquad\qquad\d\y_2=\e\bam\x^a\y_b$$
and $R=d\o$ and $T^a$  are the field strengths corresponding to Lorentz
rotations and translations respectively. The first two terms in the lagrangian
(13) are invariant under these \tran, while the potential terms are not.
As we shall see, the full action is in fact invariant under a non-linear
generalization of the \poi group.

In first order formalism, the field equations are given by:
$$\eqalign{&T^a+\gi\y^a\e_{bc}e^be^c=0\cr
&d\o+\Lq\y_2^\hm\e\ab e^ae^b=0\cr
&d\y_a+\y_b\e\bam\o+\left(\Lh\y_2^h+\gi\y_c\y^c\right)\e\ab e^b=0\cr
&d\y_2+\y_a\e\abm\e^b=0\cr}\eqno(14)$$
whose static solutions can be written in terms of the function $f$ defined
above as:
$$\e^0_t=e^1_x=\sqrt{Ef'e^f}\qquad\qquad$$
$$\o_t=\Lh E\g^\hm f^he^f-\ha f'\qquad\qquad\o_x=0$$
$$\y_0=\g\sqrt{f'\over Ee^f}\qquad\qquad\y_1=0\qquad\qquad\y_2=\g f$$

\section{5. The hamiltonian formalism}
Another advantage of the first order formalism is that it leads naturally
to a hamiltonian formulation of the model, and hence permits a
straightforward discussion of its symmetries and quantization.
In fact, after integration by parts, the lagrangian
density can be written as:
$$\eqalign{\ha\lie=&\y_a\dot e^a_x+\y_2\dot\o_x\cr
+&e^a_t(\y_a'+\e\bam\y_b\o_x+\Lh\y_2^h\e\ab e_x^b+\gi \y_c\y^c\e\ab e_x^b)
+\o_t(\y_2'+\y_a\e\abm e^b_x)\cr}\eqno(15)$$
where a dot denotes time derivative and a prime spatial derivative.

The lagrangian (15) has a canonical structure, with coordinates $(e^a_x,
\o_x)$, conjugate momenta $(\y_a,\y_2)$ and Lagrange multipliers $(e^a_t,
\o_t)$ enforcing the constraints:
$$G_a=\y_a'+\e\bam\y_b\o_x+\left(\Lh\y_2^h+\gi \y_c\y^c\right)\e\ab e_x^b=0$$
$$G_2=\y_2'+\y_a\e\ab e^b_x=0\eqno(16)$$
Combining the two constraints (16), one can deduce that
$$\ha(\y_a\y^a)'-\Lh\y_2^h\y_2'-\gi\y_a\y^a\y_2'=0\eqno(17)$$
which implies the existence of the conserved quantity
$$Q\id\y_a\y^ae^{-2\y_2/\g}-\L\left({\g\over 2}\right)^\hp\G\left(h+1,
{2\y_2\over\g}\right)\eqno(18)$$
with $\G$ the incomplete gamma function.

The study of the algebra of constraints permits to discuss the symmetries
of the theory. The calculation of the Poisson brackets of the constraints
yields
$$\{G_a,G_2\}=\e\bam G_b\qquad \{G_a,G_b\}=\e\ab\Lh h\y_2^{h-1}G_2+\gi\y_cG_c
\eqno(19)$$
with coordinate dependent structure functions.
This algebra acts locally on the fields by  the infinitesimal \tran:
$$\d e^a=d\x^a+\e\abm(\x^b\o-\x^2e^b)-\gi\e_{bc}\x^be^c\y^a\qquad\qquad
\d\o=d\x^2-\Lh h\y_2^\hm\e\ab\x^a e^b$$
$$\d\y_a=\e\bam\left[\x^2\y_b+\x_b\left(\Lh\y_2^h+\gi\y^a\y_a\right)\right]
\qquad\qquad\d\y_2=\e\bam\x^a\y_b$$
as can be checked by computing the commutators $\d e^a=\{G_a,e^b\}$, etc.

The lagrangian (13) is invariant under these transformations up to a total
derivative. Our model can therefore be considered as
a gauge theory of the group generated by the non-linear algebra (19),
realized by means of its action on the lagrangian (13).
The generalization of the usual gauge theories to non-linear algebras has
been introduced in [14], where also the special case $h=2$ of our model has
been examined.

It must be noticed, however, that in this form the algebra fails to close.
In order to construct a closed algebra one has to include in it also the
fields $\y_i$ ($i=0,1,2)$  and to consider the family of generators
$A(\y_i)+B(\y_i)G_i$, with $A$, $B$ analytic functions of $\y_i$ [11].
One has then:
$$\{\y_a,\y_2\}=\{\y_a,\y_b\}=0$$
$$\{G_2,\y_2\}=0\qquad\qquad\{G_a,\y_b\}=\e\ab\left(\Lh\y_2^h+\gi\y_c\y^c
\right)\eqno(20)$$
$$\{G_2,\y_a\}=-\{G_a,\y_2\}=\e\bam\y_b$$
The resulting algebra is a nonlinear deformation of $iso(1,2)$ of the kind
discussed in [17].

\section{6. Dirac quantization}
The model can now be quantized in the Dirac formalism, by replacing the
Poisson brackets with commutators and imposing the Gauss law on the physical
states. In a momentum representation for the wave functional,
$e^a\to i{d\over d\y_a}$, $\o\to i{d\over d\y_2}$, the constraint equations
become:
$$\left[\y_a'+i\e\bam\y_b{\de\over\de\y_2}+i\e\ab\left(\Lq\y_2^h+{1\over 2\g}
\y_c\y^c\right){\de\over\de\y_b}\right]\Q(\y_a,\y_2)=0\eqno(21)$$
$$\left(\y_2'+i\e\bam\y_a{\de\over\de\y_b}\right)\Q(\y_a,\y_2)=0
\eqno(22)$$
The solution of these equations can be written as:
$$\Q=\d(Q')e^{i\O}\q(Q)\eqno(23)$$
where $Q$ is given in (18) and
$$\O=\int{\e\ba\y_2\y_ad\y_b\over\y^c\y_c}\eqno(24)$$
The parameters $\L$ and $\g$ enter in (23) only through the parameter $Q$,
which classifies the quantum states. Some special
cases of the solution (23) have been obtained in [5,13,18].

It should be pointed out, however, that one can not straightforwardly
define a Schr\"od- inger
equation, since due to the constraints (21,22), the hamiltonian vanishes
on the physical states. This is a well-known problem in the hamiltonian
quantization of gravity and can be solved by fixing a gauge: in a
two-dimensional context it has been treated in ref. [13].

\section{7. Conclusions}
We have shown that most of the results obtained in two-dimensional gravity
with quadratic curvature and torsion can be extended to the case of an action
containing arbitrary powers of the curvature scalar. A possible generalization
of these results would be to consider actions containing arbitrary functions of
of the curvature, which can be treated essentially by the same methods used
here [3,4].
Another interesting point would be to find the most general solutions of the
field equations, including time-dependent ones, which we have not considered.
It seems plausible that this can be achieved by means of a suitable
generalization of the procedure followed in ref. [9] and [10] in the special
case of quadratic curvature.

\beginref
\ref [1] H.J. Schmidt, \JMP{32}, 1562 (1991);
\ref [2] S. Mignemi, \PR{D50}, 4733 (1994);
\ref [3] S.N. Solodukhin, \PR{D51}, 591 (1995);
\ref [4] S. Mignemi and H.-J. Schmidt, \CQG{12}, 845 (1995);
\ref [5] S. Mignemi, in press on Ann. Phys.;
\ref [6] C. Teitelboim, in {\sl Quantum Theory of gravity}, S.M. Christensen,
 ed. (Adam Hilger, Bristol, 1984); R. Jackiw, {\sl ibidem};
\ref [7] G. Mandal, A.M. Sengupta and S.R. Wadia, \MPL{A6}, 1685 (1991);
\ref [8] D. Cangemi and R. Jackiw, \PRL{69}, 233 (1992);
R. Jackiw, Theor. Math. Phys. {\bf 9}, 404 (1992);
\ref [9] M.O. Katanaev and I.V. Volovich, \PL{B175}, 413 (1986); \AP{N.Y.}
{197}, 1 (1990); M.O. Katanaev, \JMP{31}, 882 (1990); {\bf 32}, 2483 (1991);
\ref [10] W. Kummer and D.J. Schwarz, \PR{D45}, 3628 (1992);
\ref [11] H.Grosse, W. Kummer, P. Pre\v snajder and D.J. Schwarz,
\JMP{33}, 3892 (1992);
\ref [12] T. Strobl, \IJMP{A8}, 1383 (1993);
\ref [13] T. Strobl, \IJMP{D3}, 281 (1994);
\ref [14] N. Ikeda, \AP{N.Y.}{235}, 435 (1994); N. Ikeda and K.I. Izawa,
\PTP{89},
223 (1993); {\bf 89}, 1077 (1993); {\bf 90}, 237 (1993);
\ref [15] K.G. Akdeniz, A. Kizilers\"u and A. Rizao\v glu, \PL{B215}, 81
(1988);
K.G. Akdeniz, \"O.F. Dayi and A. Kizilers\"u, \MPL{A7}, 1757 (1992);
\ref [16] T. Fukuyama and K. Kamimura, \PL{B160}, 259 (1985);
K. Isler and C.A. Trugenberger, \PRL{63}, 834 (1989);
A.H. Chamseddine and D. Wyler, \PL{B228}, 75 (1989);
\ref [17] K. Schoutens, A. Sevrin and P. van Nieuwenhuizen, \CMP{124}, 87
(1989);
\ref [18] D. Cangemi and R. Jackiw, \PR{D50}, 3916 (1994).
\endref
\end